\documentclass[prd,preprint,superscriptaddress,amsmath,amssymb,nofootinbib]{revtex4}
\usepackage{graphicx}% Include figure files
\usepackage{dcolumn}% Align table columns on decimal point
\usepackage{bm}% bold math
\usepackage{amssymb}
\usepackage{amsmath}
\usepackage{epsfig}
\usepackage{color}
\usepackage{slashed}
\usepackage{hhline}
\usepackage{bbm}
\usepackage{hyperref}
\usepackage{tikz-feynman}
\tikzfeynmanset{compat=1.1.1}

\def\be{\begin{equation}}
\def\ee{\end{equation}}
\newcommand{\bea}{\begin{eqnarray}}
\newcommand{\eea}{\end{eqnarray}}

\newcommand{\mR}{m_{\eta_R}}
\newcommand{\mI}{m_{\eta_I}}
\newcommand{\mC}{m_{\eta^{\pm}}}

\newcommand{\gev}{\,\mathrm{GeV}}
\newcommand{\kev}{\,\mathrm{keV}}
\newcommand{\tev}{\,\mathrm{TeV}}

\newcommand{\br}{\mathrm{BR}}

\begin{document}

%%%%%%%%%
%\title{Self-Consistent Framework for Neutrino Masses: Cut-off Scale from RGE Flow and Loop-Induced VEVs under Non-Invertible Symmetry}
\title{Can a minimal radiative seesaw explain the LZ 248 keV event?}

\author{Hiroshi Okada}
\email{hiroshi3okada@htu.edu.cn}
\affiliation{Department of Physics, Henan Normal University, Xinxiang 453007, China}

\author{Yoshihiro Shigekami}
\email{shigekami@htu.edu.cn}
\affiliation{Department of Physics, Henan Normal University, Xinxiang 453007, China}

\author{Jia-Jun Wu}
\email{wujiajun@htu.edu.cn}
\affiliation{Department of Physics, Henan Normal University, Xinxiang 453007, China}

\date{September 2026}

\begin{abstract}
We interpret the recently reported 248~keV nuclear recoil event in the LUX-ZEPLIN (LZ) experiment via inelastic dark matter scattering within the minimal Scotogenic model. A sub-MeV mass splitting between neutral inert scalars suppresses low-energy scattering while permitting signals from the high-velocity halo tail. Crucially, co-annihilation with nearly degenerate right-handed fermions accommodates the thermal relic density for dark matter masses up to $\sim {\cal O} (1)$~TeV, 
%$\sim 1.5$~TeV, 
extending the viable range significantly beyond the pure inert doublet model limit while evading direct detection bounds. However, to resolve the severe tension with IceCube neutrino limits on solar capture, we extend this minimal framework by introducing a hidden $U(1)_X$ gauge symmetry that naturally leads us to tiny $\lambda_5$ coupling at the one-loop level. This realizes an isospin-violating scenario that suppresses dark matter capture in the Sun while preserving the coherent scattering signal in the Xenon-based LZ detector. We numerically verify that this extended framework naturally generates neutrino masses and satisfies constraints from Big Bang Nucleosynthesis and indirect detection, providing a robust and testable solution to the LZ anomaly.
\end{abstract}

\maketitle

\newpage

\section{Introduction}
\label{sec:intro}

The elucidation of the origins of neutrino masses and dark matter (DM) stands as one of the most compelling and fundamental challenges in particle physics and cosmology, strongly motivating the exploration of physics beyond the Standard Model (BSM). 
A framework that plays a pivotal role in addressing these mysteries is the minimal radiative seesaw model, originally proposed by Ma~\cite{Ma:2006km}, commonly known as the Scotogenic model. 
The profound appeal of this model lies in the intimate and direct connection it establishes between neutrinos and DM. 
The smallness of the neutrino masses is naturally interpreted as a consequence of their generation occurring solely at the loop level, mediated by interactions with the dark sector. 
Concretely, the model requires only the addition of a single inert scalar doublet ($\eta$) and three heavy right-handed Majorana fermions ($N_R$), all of which are odd under an exact $Z_2$ symmetry while all Standard Model (SM) fields remain even. 
This remarkably simple and elegant field content naturally provides viable DM candidates: either the neutral component of the inert doublet or the lightest of the right-handed neutrinos. 

In this paper, we focus on the scalar DM branch, specifically the neutral component of the inert doublet, $\eta^0 = (\eta_R + i \eta_I) / \sqrt{2}$. 
We propose that the recently reported 248~keV high-energy nuclear recoil event in the LUX-ZEPLIN (LZ) experiment~\cite{LZ:2026axp} can be naturally explained via inelastic dark matter scattering within this framework.\footnote{Several interpretations of the LZ anomaly have been discussed in Refs.~\cite{Su:2026rwz,Fan:2026kxx,Freese:2026sga,Wu:2026nhi,Lou:2026idn,Yin:2026jnn,Nomura:2026qyq,DiMauro:2026ldr,Pospelov:2026ewn,Visinelli:2026kgt,Yamashita:2026ump,Chattopadhyay:2026ryw,Smirnov:2026aqk,Du:2026guj,Rodd:2026tyn,McCabe:2026crm,Jeesun:2026vzo,Unwin:2026rdp,Dent:2026bji,deLima:2026shq,Gu:2026vto,Baer:2026fpy,Lee:2026wof,Wang:2026ytg,Yang:2026wpb,Kotlarski:2026pep,Liang:2026coz,DiMauro:2026dqp,Das:2026uyy,Alhazmi:2026efz,Okada:2026eol,Ahmed:2026qjg,Du:2026lpa,Bandyopadhyay:2026gjw,Kannike:2026qyl,Borah:2026zwf,Bose:2026ndd,Bisal:2026khf,Cheung:2026byg,Egorov:2026dpr,Bamwidhi:2026vdu,Yuan:2026djt,Elahi:2026vlm,Zhu:2026dag,Aghaie:2026vsu,Asadi:2026iot,Lee:2026xxh,Lee:2026jxl,Khan:2026nwp,Langhoff:2026ujr,Chatterjee:2026scv,He:2026hqz,Fan:2026hzw,Chattaraj:2026fxn,Qi:2026vyp,Li:2026fci,Kumar:2026lgi,Frolovsky:2026tvq,Heikinheimo:2026kwp,Nguyen:2026lui}.} 
The sub-MeV mass splitting between the real ($\eta_R$) and imaginary ($\eta_I$) neutral components plays a crucial role. 
This splitting kinematically suppresses ordinary elastic scattering at low energies, thereby evading stringent direct detection bounds, while simultaneously permitting rare, high-energy signals originating from the high-velocity tail of the dark matter halo distribution. 
This mechanism provides an optimal and highly motivated framework for interpreting the LZ anomaly. 

A critical advantage of the Scotogenic model over the pure Inert Doublet Model (IDM) lies in the thermal history of the dark sector, specifically the mechanism for accommodating the observed thermal relic density. 
In the pure IDM, achieving the correct relic density in the heavy mass regime ($m_{\text{DM}} \gtrsim 500$~GeV) necessitates large scalar quartic couplings to enhance the annihilation cross section into gauge bosons and Higgs particles~\cite{Hambye:2009pw}. 
However, such large couplings inevitably induce large Higgs portal elastic scattering cross sections at tree level, which are severely constrained or entirely excluded by current direct detection experiments. 
In stark contrast, the Scotogenic model introduces the right-handed neutrinos $N_R$, which can participate in the thermal freeze-out via co-annihilation processes. 
This co-annihilation effect effectively reduces the required annihilation cross section of the scalar DM, thereby significantly enhancing the relic density for a given mass. 
Consequently, one can successfully fit the observed relic density with small Higgs portal couplings even for heavy dark matter masses up to $\sim {\cal O} (1)$~TeV. 
%$\sim 1.5$~TeV. 
This mechanism effectively evades the stringent direct detection limits that plague the pure IDM scenario, vastly expanding the viable parameter space capable of explaining the LZ high-recoil event. 

Furthermore, we perform an extensive numerical scan within the minimal Ma model and successfully identify viable parameter spaces that simultaneously accommodate the LZ 248~keV anomaly, the observed thermal relic density, and neutrino oscillation data. As is well known, this framework naturally incorporates the standard phenomenological predictions of the Scotogenic model, such as charged-lepton flavor violations (e.g., $\mu \to e\gamma$) and compressed scalar spectra testable at colliders; thus, we briefly summarize them without elaborating on the well-established details. Crucially, however, we reveal that the minimal setup encounters a severe tension with the IceCube neutrino observatory limits regarding solar capture. To resolve this conflict, we extend the model by introducing a hidden $U(1)_X$ gauge symmetry that naturally leads us to tiny $\lambda_5$ coupling at the one-loop level, realizing an isospin-violating scenario that suppresses dark matter capture in the Sun while preserving the coherent scattering signal in the Xenon-based LZ detector. Finally, we comprehensively evaluate other experimental and cosmological constraints, including Big Bang Nucleosynthesis (BBN), the Cosmic Microwave Background (CMB), and indirect detection limits from Fermi-LAT, demonstrating that the proposed extended framework remains robust and consistent with all current phenomenological bounds.

The paper is organized as follows. 
After reviewing the model setup and flavor constraints in Sec.~\ref{sec:model}, we discuss the relic density and the kinematic interpretation of the LZ event in Secs.~\ref{sec:thermal} and \ref{sec:signal}, respectively. 
We present our numerical results in Sec.~\ref{sec:results}, address the IceCube tension via a hidden $U(1)_X$ extension in Sec.~\ref{sec:IceCube}. 
Finally, we conclude our discussion in Sec.~\ref{sec:conclusion}. 
%, and conclude in Sec.~VII.

\section{Model setup, neutrino mass matrix, and flavor physics}
\label{sec:model}

In this section, we briefly review the original Sctogenic model we focus on in the current study.

\subsection{Scalar spectrum and the inelastic splitting}
\label{sec:MaModel}

The Ma model requires some additional fields: an inert doublet $\eta \sim (\mathbf{2}, 1/2)$ and three singlet Majorana fermions $N_i \sim (\mathbf{1}, 0)$ under $SU(2)_L \times U(1)_Y$ gauge symmetries. 
They are odd charges under the $\mathbb{Z}_2$ that assures the stability of DM candidate, while all the SM particles are even charges. 
The new particles are denoted by
\begin{equation}
\eta = \begin{pmatrix}
\eta^+ \\
(\eta_R + i \eta_I) / \sqrt{2}
\end{pmatrix} \, , 
%\qquad \langle \eta \rangle = 0 \, , 
\qquad N_{R_i} \, , 
\end{equation}
where a vacuum expectation value (VEV) of $\eta$ is $\langle \eta \rangle = 0$ as it is the inert doublet. 
The relevant Lagrangian for the lepton sector is
\begin{equation}
\mathcal{L} = - f^{\ell}_{\alpha} \overline{L_{L_{\alpha}}} H \ell_{R_{\alpha}} - Y_{\alpha i} \overline{L_{L_{\alpha}}} \widetilde{\eta} N_{R_i} + \mathrm{h.c.} - \frac{1}{2} \sum_i M_i \overline{N^C_{R_i}} N_{R_i} \, ,
\label{eq:yukawa}
\end{equation}
where $L_L$ and $\ell_R$ are the SM lepton doublet and singlet, respectively, and $H$ denotes the SM Higgs doublet. 
The first term provides the charged-lepton mass eigenvalues after the spontaneous symmetry breaking of the SM Higgs that is denoted by $\langle H \rangle = [0, v / \sqrt{2}]^T$, and $\widetilde{\eta} \equiv i \sigma_2 \eta^*$ with $\sigma_2$ being the second Pauli matrix.
Without loss of generality, the term $f$ and Majorana mass term $M_i$ can be diagonal.
The relevant Higgs potential is given by
\begin{align}
V = &- \mu_1^2 H^{\dagger} H + \mu_2^2 \eta^{\dagger} \eta + \lambda_1 (H^{\dagger} H)^2 + \lambda_2 (\eta^{\dagger} \eta)^2 \nonumber \\
&+ \lambda_3 (H^{\dagger} H)(\eta^{\dagger} \eta) + \lambda_4|H^{\dagger} \eta|^2 + \frac{\lambda_5}{2} \big[ (H^{\dagger} \eta)^2 + \mathrm{h.c.} \big] \, .
\label{eq:potential}
\end{align}
After the spontaneous symmetry breaking of the SM Higgs, each mass eigenvalue is given by
\begin{align}
\mC^2 &= \mu_2^2 + \frac{1}{2} \lambda_3 v^2 \, , &\mR^2 &= \mu_2^2 + \lambda_L v^2 \, , \nonumber \\
\mI^2 &= \mu_2^2 + \frac{1}{2} (\lambda_3 + \lambda_4 - \lambda_5) v^2 \, , &\lambda_L &= \frac{1}{2} (\lambda_3 + \lambda_4 + \lambda_5) \, .
\label{eq:masses}
\end{align}
Here, we choose $\eta_R$ as DM, therefore $M_i, \mI, \mC > \mR$ should be satisfied. 
Defining $\delta \equiv \mI - \mR > 0$ and $\Delta_{\pm} \equiv \mC - \mR > 0$, one obtains
\begin{equation}
\lambda_5 = - \frac{2 \mR \delta + \delta^2}{v^2} \simeq - \frac{2 \mR \delta}{v^2} \, .
\label{eq:lambda5}
\end{equation}
Thus $\mR \sim 1 \tev$ and $\delta \sim 0.37 \, \mathrm{MeV}$ correspond to $|\lambda_5| \sim 1.2 \times 10^{-5}$. 
%%%%%%%%%
%A lepton-number assignment $L(L_{\alpha})=1$, $L(\eta)=-1$, $L(N_i)=0$ makes the Yukawa and Majorana mass terms invariant; $\lambda_5$ then breaks this symmetry by two units. The limit $\lambda_5\to0$ restores the perturbative lepton-number symmetry and removes the radiative Majorana neutrino mass. Small $\lambda_5$ is therefore technically natural within this theory.
%%%%%%%%%

\subsection{Neutrino mass matrix and flavor observables}
\label{sec:PhenoMaModel}

The neutrino mass matrix is simply given by
\begin{align}
m_{\nu} &= Y \Lambda Y^T \, , \label{eq:mnu} \\
\Lambda_i &= \frac{M_i}{32 \pi^2} \left[ \frac{\mR^2}{\mR^2 - M_i^2} \ln \frac{\mR^2}{M_i^2} - \frac{\mI^2}{\mI^2 - M_i^2} \ln \frac{\mI^2}{M_i^2} \right] \, . \label{eq:kernel}
\end{align}
Then, the neutrino mass matrix is diagonalized by $D_{\nu} = U^T m_{\nu} U$, where $U$ is the observed mixing matrix including two Majorana phases. 
It can lead us to the Casas--Ibarra parametrization~\cite{Casas:2001sr}:
\begin{equation}
Y = U^* \sqrt{D_{\nu}} \, R \Lambda^{-1/2} \, , \qquad R R^T = \mathbf{I} \, ,
\label{eq:ci}
\end{equation}
where the matrix $R$ is three by three complex orthogonal matrix. 
NuFIT~6.1~\cite{Esteban:2024eli}\footnote{See \href{http://www.nu-fit.org/}{http://www.nu-fit.org/} for latest results.} provides the latest experimental results both for normal hierarchy (NH) and inverted hierarchy (IH). 
Thus, we can numerically determine the Yukawa coupling $Y$ via the experimental results and several free parameters in the Ma model. 
Note that $Y \lesssim \sqrt{4 \pi}$ should be satisfied as the perturbative limit. 

For reference, the dipole contribution to $\mu \to e \gamma$ can be found in Ref.~\cite{Toma:2013zsa}. 
The formulated $Y$ in the neutrino sector can be conveyed to the cLFVs. 
Even though cLFVs depend on flavor indices of $Y$, we simply focus on the most stringent process of $\mu \to e \gamma$ whose branching ratio should be less than $1.5 \times 10^{-13}$ at 90\% CL from MEG~II~\cite{MEGII:2025gzr}. 
The branching ratio in the current model is given by
\begin{align}
&\br (\mu \to e \gamma) \simeq \frac{3 (4 \pi)^3 \alpha_{\rm em}}{4 G_F^2} |A_D|^2 \, , \\
&A_D = \sum_i \frac{Y_{e i}^* Y_{\mu i}}{2 (4 \pi)^2 \mC^2} F_2 (x_i) \, , \quad F_2 (x) = \frac{1 - 6 x + 3 x^2 + 2 x^3 - 6 x^2 \ln x}{6 (1 - x)^4} \, ,
\label{eq:lfv}
\end{align}
where $x_i = \frac{M_i^2}{\mC^2}$ and $\alpha_{\rm em} \approx 1 / 137$ is the fine structure constant. 
In the numerical analysis, we adopt a compressed charged-neutral scalar spectrum with $m_{\eta^{\pm}} \simeq m_{\eta_I}$; the electroweak precision constraints are evaluated explicitly with 2HDMC. 
The one-loop $\mu \to 3 e$ and coherent $\mu$--$e$ conversion expressions are evaluated by Ref.~\cite{Toma:2013zsa}. 
The corresponding experimental thresholds are specified in Sec.~\ref{sec:inputs}.

\section{Relic density with co-annihilation}
\label{sec:thermal}

As mentioned in the Introduction, this Scotogenic model has a DM candidate, and we focus on the case of the $\eta_R$ DM. 
For our purpose, we include co-annihilation process for the calculation of its relic density. 
%\subsection{Chemical equilibrium and effective annihilation}
%When conversions maintain relative chemical equilibrium among the odd particles, their total density can be evolved with a single coannihilation equation~\cite{Griest:1990kh}. 

Here, we formulate our main (co-)annihilation processes; specifically the interactions among the DM and quasi-degenerated three right-handed Majorana neutrinos $N_{R_i}$ in the nonrelativistic Maxwell--Boltzmann approximation. %, where $\eta_R$ is regarded as the DM candidate. 
The thermally averaged cross section is
\begin{align}
\langle \sigma_{\rm eff} v \rangle &= \sum_{a, b} r_a (x) r_b (x) \langle \sigma_{a b} v \rangle \, , \nonumber \\
r_a (x) &= \frac{g_a (1 + \Delta_a)^{3/2} e^{- x \Delta_a}}{g_{\rm eff} (x)} \, , \qquad g_{\rm eff} (x) = \sum_a g_a (1 + \Delta_a)^{3/2} e^{- x \Delta_a} \, ,
\label{eq:coann}
\end{align}
where $x = \mR / T$, $\Delta_a = (m_a - \mR) / \mR$, and $g_a$ is internal degrees of freedom for each the degenerate particle. 
%Each real neutral scalar has one internal degree of freedom; $\eta^+$ and $\eta^-$ contribute one each, and each Majorana singlet contributes two.

%If singlet annihilation channels are weak, adding thermally populated singlets increases the denominator in Eq.~\ref{eq:coann} more than the annihilation numerator. For approximately degenerate scalars and singlets, and otherwise unchanged scalar annihilation, this gives the estimate
%Since the recent direct detection bounds are so strict, we do not rely on any interactions via the Higgs potential to explain the relic density.
Assuming that the annihilation channels involving the thermally populated singlet fermions are weaker than the dominant scalar annihilation channels, we can approximately evaluate the ratio of relic density between only annihilations and annihilations in addition to co-annihilations with $N_R$ as follows:
\begin{equation}
\frac{\Omega_{\eta_R} h^2}{\Omega_{\rm IDM} h^2} \sim \left( \frac{g_{\eta} + 2 n_N}{g_{\eta}} \right)^2 \, , \qquad g_{\eta} \simeq 4 \, ,
\label{eq:dilution}
\end{equation}
that is the fermion-coannihilation mechanism discussed in Ref.~\cite{Klasen:2013jpa}. 
Here, $n_N$ denotes the number of singlet Majorana fermions whose masses are close to $m_{\eta_R}$.
In fact, the relic density can be enhanced up to 6.25 with three degenerated $N_R$, which suggests that the allowed DM mass range can be relaxed up to $\sim {\cal O} (1)$~TeV 
%$1.5$ TeV 
without relying on interactions via Higgs potential. Thus, this mechanism is crucially important to explain the LZ event without conflict of the latest severe direct detection bounds as we will show later. 

%The estimates for one, two, and three singlets are 2.25, 4, and 6.25. Finite mass gaps, suppressed charged-scalar populations, additional annihilation channels, and a shifted freeze-out temperature modify these values; they are not strict bounds. 

Before applying the standard co-annihilation formalism~\cite{Griest:1990kh}, it is crucial to ensure that the conversion processes (e.g., $\eta_R + \ell \leftrightarrow N_i + \dots$) are sufficiently rapid to maintain chemical equilibrium between the scalar DM and the right-handed neutrinos throughout the freeze-out epoch. 
Given the small Yukawa couplings required to satisfy cLFV bounds, there is a potential risk of early decoupling of the singlet fermions. 
We have explicitly verified that, for all our benchmark points, the conversion rates significantly exceed the Hubble expansion rate ($\Gamma_{\text{conv}} / H \gg 1$) at the freeze-out temperature. 
Thus, the assumption of relative chemical equilibrium is well justified, and the co-annihilation enhancement of the relic density is robust. 

{
The abundance window is given by $0.120 \pm 0.001$~\cite{Planck:2018vyg} at 1$\sigma$ confidence level. 
In our numerical analysis, however, we take the window up to 3$\sigma$ confidence level;
\begin{equation}
0.117 \le \Omega_{\eta_R} h^2 \le 0.123 \, .
%0.11\le\Omega_{\eta_R}h^2\le0.13.
\label{eq:relicwindow}
\end{equation}
}

\section{Direct detections}
\label{sec:signal}

\subsection{LZ: Kinematic interplay between inelastic signal}
\label{sec:kinematics_dd}

A crucial aspect of interpreting the LZ high-recoil event within the Scotogenic model is the distinct kinematic roles of inelastic and elastic scattering. 
While the former generates the signal, the latter is subject to stringent direct detection (DD) constraints.
%A crucial aspect of interpreting the LZ high-recoil event within the Scotogenic model lies in the fundamentally different kinematic roles played by the inelastic scattering, generates the signal, and the elastic scattering, which is subject to stringent direct detection (DD) bounds.

The inelastic process $\eta_R + N \to \eta_I + N$, mediated by the $Z$-boson exchange, possesses a relatively large cross-section due to the unsuppressed electroweak gauge coupling. 
However, the sub-MeV mass splitting $\delta = m_{\eta_I} - m_{\eta_R} \simeq 360$~keV introduces a significant kinematic threshold. 
The minimum incident velocity required to produce a nuclear recoil of energy $E_R$ is given by
\begin{equation}
v_{\min} (E_R) = \frac{1}{\sqrt{2 m_N E_R}} \left( \frac{m_N E_R}{\mu_{\chi N}} + \delta \right) \, ,
\end{equation}
where $m_N$ is the target nucleus mass and $\mu_{\chi N}$ is the dark matter-nucleus reduced mass. 
For $\delta \simeq 360$~keV and $E_R \sim \mathcal{O}(10\text{--}100)$~keV, the required $v_{\min}$ significantly exceeds the characteristic velocity of the dark matter halo ($v_0 \simeq 238$~km/s). 
Consequently, the inelastic scattering rate at low and intermediate recoil energies is kinematically forbidden, and the observable signal is entirely restricted to the extreme high-velocity tail of the halo distribution. 
This mechanism naturally allows the model to generate a rare 248~keV event while completely evading the low-energy bounds.

The coherent weak-charge scattering cross-section is
\begin{equation}
\frac{d \sigma_A}{d E_R} = \frac{m_A \sigma_A^0}{2 \mu_A^2 v^2} F_A^2 (E_R) \, ,
\end{equation}
where $\sigma_A^0 = \frac{G_F^2 \mu_A^2}{2 \pi} Q_W^2$, the weak charge $Q_W = (A - Z) - (1 - 4 \sin^2 \theta_W) Z$ with the weak mixing angle $\theta_W$, and $F_A (E_R)$ is the Helm form factor. %citation needed?
The differential event rate per unit detector mass is then calculated as
\begin{equation}
\frac{d R}{d E_R} = \frac{\rho_{\chi}}{m_{\eta_R}} \sum_A \frac{\xi_A}{m_A} \int_{v \geq v_{\min} (E_R)} d^3 v \, f_{\text{lab}} (v) \, v \, \frac{d \sigma_A}{d E_R} \, ,
\end{equation}
where $\xi_A$ denotes the isotope mass fraction of natural Xenon. 
We adopt a standard truncated Maxwellian halo distribution with local density $\rho_{\chi} = 0.3$~GeV cm$^{-3}$, characteristic velocity $v_0 = 238$~km/s, escape velocity $v_{\text{esc}} = 544$~km/s, and a fixed laboratory velocity $v_{\text{lab}} = 254$~km/s as a reference scenario.

\subsection{Direct detection bounds for elastic direct detections}
\label{sec:dd}

In stark contrast, the elastic scattering $\eta_R + N \to \eta_R + N$, which is mediated by the Higgs portal coupling $\lambda_L$ at tree level and by electroweak loop corrections, has no such kinematic threshold ($\delta = 0$). 
The minimum velocity $v_{\min}$ for elastic scattering starts from nearly zero, meaning that the entire bulk of the dark matter halo velocity distribution contributes to the scattering rate. 
Even if the elastic cross-section is intrinsically smaller than the inelastic one, the integration over the full velocity range results in an overwhelming number of low-energy recoil events. 

This dichotomy dictates the strategy for model building. 
The LZ 248~keV event can be accommodated by the inelastic channel without conflicting with low-energy DD limits, provided that the elastic scattering channel is sufficiently suppressed. 
In the pure IDM, accommodating the thermal relic density for heavy dark matter masses ($m_{\eta_R} \gtrsim 500$~GeV) requires large scalar quartic couplings, which in turn induce large Higgs portal elastic cross-sections that are severely constrained by current DD experiments. 
In the Scotogenic model, however, the co-annihilation with right-handed neutrinos enhances the relic density, allowing us to keep the Higgs portal coupling $\lambda_L$ small. 
Furthermore, the remaining tree-level and loop-level elastic contributions can be further suppressed or canceled, ensuring that the model safely evades the stringent DD bounds while explaining the LZ anomaly. 

The minimal Scotogenic model introduces the Yukawa couplings $Y_{\alpha i}$ connecting the inert doublet $\eta$, the lepton doublets $L_{\alpha}$, and the right-handed neutrinos $N_i$. 
Since these couplings are responsible for the radiative neutrino masses and the fermion co-annihilation discussed in this work, a natural concern is whether they induce large one-loop corrections to the elastic direct-detection cross section, potentially competing with or overwhelming the dominant inert-doublet gauge loop contributions. 
In fact, we have one-loop triangle contributions via $N_R$ and $\nu_L$, and hence, it connects to quarks through $\nu_L$ interaction with the $Z$ boson. 
%that connects to quarks. 
In case of $Z$ boson mediation, the scattering cross section is typically $10^{-53}$~cm$^2$ under the following fixed values $m_{\eta_R} \sim m_{\eta^{\pm}} \sim M_i = 1 \tev$ and  $Y \sim 10^{-3}$. 
Thus, we can totally ignore the loop contributions.

\section{Numerical results}
\label{sec:results}

\subsection{Scalar and singlet spectra}
\label{sec:inputs}

We present numerical results for the minimal Scotogenic model, with the relic density evaluated using micrOMEGAs~7.1.4~\cite{Belanger:2026asz}. 
The points satisfy the neutrino and charged-lepton flavor constraints, scalar-sector theoretical and electroweak precision constraints checked with 2HDMC~\cite{Eriksson:2009ws}, and the elastic direct-detection and indirect-detection criteria used in the scan. 
Of the 113 candidates, 34 satisfy the relic-density requirement $0.117 \leq \Omega_{\eta_R} h^2 \leq 0.123$. Their masses and neutral-scalar splittings span $\mR = 691.19$--$1479.37$~GeV and $\delta = 360$--$380 \kev$.

\begin{figure}[!thbp]
\centering
\includegraphics[width=0.83\linewidth]{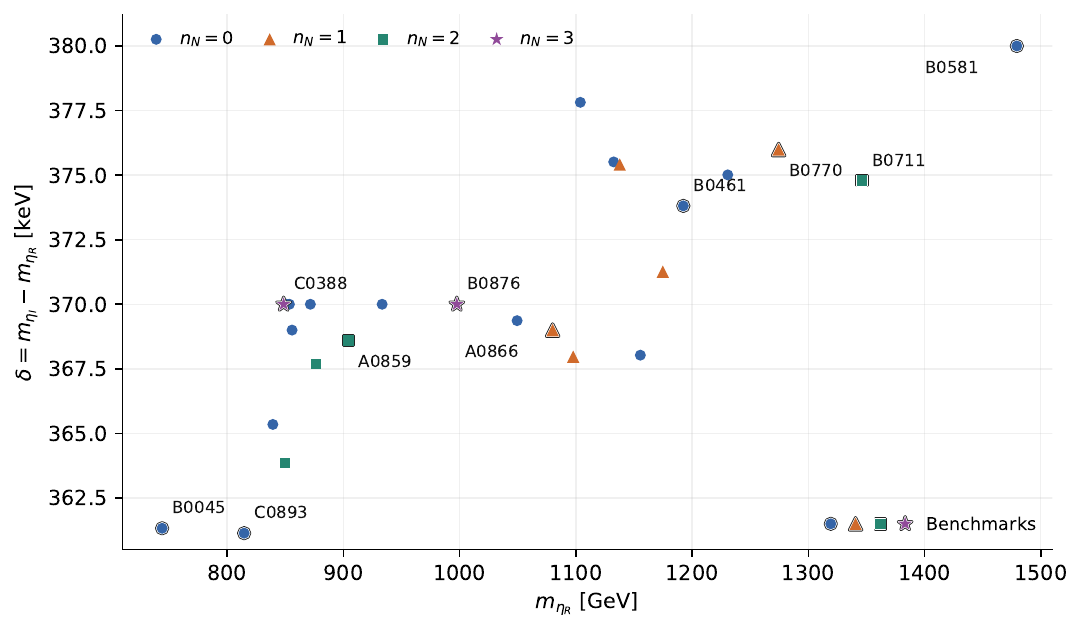}
\caption{Points satisfying the relic-density requirement in the $(\mR, \delta)$ plane. 
Colors and marker shapes distinguish the number $n_N$ of nearly degenerate singlet fermions, and black outlines identify the ten benchmarks. 
The figure displays 26 of the 34 selected points, with the $\delta=360 \kev$ row omitted. }
\label{fig:sample}
\end{figure}

Figure~\ref{fig:sample} shows the distribution of these points. 
Here $n_N = 0, 1, 2, 3$ denotes the number of singlet Majorana fermions assigned masses close to $m_{\eta_R}$ in the scan; their individual contributions at freeze-out are weighted by the corresponding Boltzmann factors. 
Ten benchmarks covering both neutrino mass orderings and the four values of $n_N$ are highlighted. 
The eight points at $\delta = 360 \kev$ are omitted from the plot but retained in the sample. 

Table~\ref{tab:bench} lists the benchmark spectra and relic densities. 
To quantify the effect of the singlet fermions, we define $r_{\Omega} = \Omega_{\eta_R} / \Omega_{\rm IDM}$, where $\Omega_{\eta_R}$ is the Scotogenic relic density, and $\Omega_{\rm IDM}$ is the abundance in the inert doublet model with the same scalar masses and couplings. 
The four $n_N = 0$ benchmarks have $r_{\Omega} \simeq 1$, while the other six give $r_{\Omega} = 2.30$--$4.78$. 
For B0876, the singlets increase $\Omega h^2$ from 0.02480 to 0.11846. 
The highest-mass benchmark, B0581, instead has an abundance essentially equal to that in the corresponding IDM calculation. 

\begin{table}[htbp]
\centering
\small
\setlength{\tabcolsep}{5pt}
\renewcommand{\arraystretch}{1.02}
\begin{tabular}{lccccc@{\hspace{12pt}}c@{\hspace{12pt}}c@{\hspace{12pt}}c}
\hline\hline
Point & Ord. & $n_N$
& $m_{\eta_R}\,[\mathrm{GeV}]$
& $\delta\,[\mathrm{keV}]$
& $\Delta_\pm\,[\mathrm{GeV}]$
& $\Omega_{\eta_R}h^2$
& $r_\Omega$
& $N_{\rm WS}$\\
\hline
B0045 & NH & 0 &  744.04 & 361.327 &  6.31 & 0.11975 & 1.000 & 1.039\\
C0893 & NH & 0 &  814.61 & 361.137 &  7.03 & 0.11796 & 1.000 & 2.682\\
C0388 & IH & 3 &  848.58 & 370.000 & 17.27 & 0.11809 & 3.454 & 0.224\\
A0859 & NH & 2 &  904.28 & 368.596 & 14.05 & 0.11909 & 2.437 & 0.726\\
B0876 & IH & 3 &  997.57 & 370.000 & 21.15 & 0.11846 & 4.777 & 1.061\\
A0866 & IH & 1 & 1080.00 & 369.000 & 13.96 & 0.11919 & 2.296 & 2.323\\
B0461 & NH & 0 & 1192.44 & 373.809 &  8.54 & 0.11776 & 1.000 & 1.130\\
B0770 & IH & 1 & 1274.45 & 375.991 & 15.56 & 0.12270 & 2.783 & 0.883\\
B0711 & NH & 2 & 1346.09 & 374.786 & 14.51 & 0.12191 & 2.427 & 1.580\\
B0581 & NH & 0 & 1479.37 & 380.000 &  8.74 & 0.12137 & 1.000 & 0.591\\
\hline\hline
\end{tabular}
\caption{Ten benchmark points. NH and IH denote the normal and inverted neutrino mass orderings, respectively, and $\Delta_\pm=m_{\eta^\pm}-m_{\eta_R}$ is the charged-scalar mass splitting. The last column gives the expected signal count $N_{\rm WS}$ after the LZ working-search selection, evaluated with Eq.~\eqref{eq:count} at $v_{\rm lab}=254$~km/s.}
\label{tab:bench}
\end{table}

\subsection{Xenon recoil spectra}
\label{sec:selected_yields}

The expected number of signal events passing the LZ working-search selection is
\begin{equation}
N_{\rm WS} = \mathcal{E} \int d E_R \, \epsilon_{\rm WS} (E_R) \frac{d R}{d E_R} \, , \qquad \mathcal{E} = 2.84~{\rm tonne\,year} \, ,
\label{eq:count}
\end{equation}
where $\mathcal{E}$ is the exposure, and $\epsilon_{\rm WS} (E_R)$ is the working-search selection efficiency at true recoil energy $E_R$. 
The rate $d R / d E_R$ is summed over the natural Xenon isotopes and expressed per unit detector mass. 
With the halo parameters specified above, including $v_{\rm lab} = 254$~km/s, the 34 points yield $N_{\rm WS} = 0.184$--$4.068$, and the ten benchmarks give $0.224$--$2.682$. 
These counts are integrated over the efficiency-weighted recoil spectrum. 

\begin{figure}[htbp]
\centering
\includegraphics[width=\linewidth]{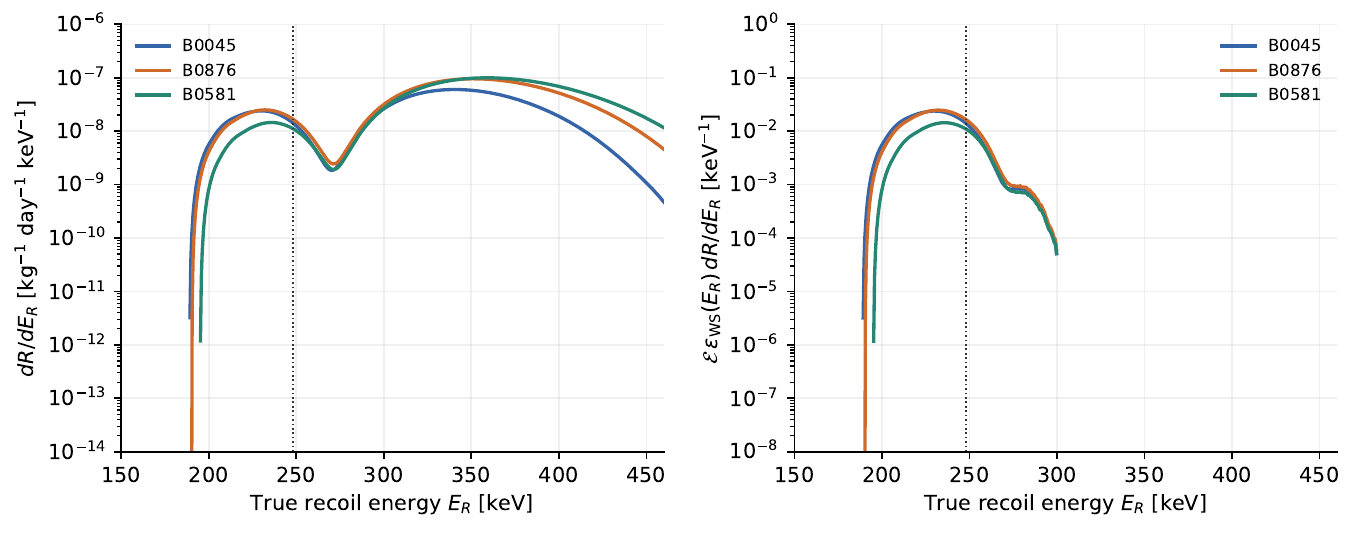}
\caption{Natural-Xenon spectra for B0045, B0876, and B0581: the differential recoil rate (left) and the expected event yield per unit recoil energy after the working-search selection (right). 
Both horizontal axes show true recoil energy; the vertical line marks 248~keV. }
\label{fig:spectra}
\end{figure}

Figure~\ref{fig:spectra} shows three benchmarks across the mass range. 
Their spectra extend through 248~keV, with total selected yields of order unity, giving the expected scale of a rare high-energy signal in LZ. 
The rate is particularly sensitive to the fastest halo particles: at 248~keV, the ten benchmarks require minimum speeds of approximately 786--797~km/s for scattering on $^{131}$Xe, close to the halo cutoff in the laboratory frame, $v_{\rm esc} + v_{\rm lab} = 798$~km/s. 
Keeping the other inputs fixed, lowering $v_{\rm lab}$ to 239~km/s reduces the integrated yields by factors of 55--585, whereas raising it to 269~km/s increases them by factors of 9.4--30.1. 
Thus, small changes in the population above the inelastic threshold have a large effect on the expected event count.

\section{Evading IceCube Constraints via a Hidden $U(1)_X$ Gauge Symmetry}
\label{sec:IceCube}

\subsection{The Tension between the LZ 248~keV Event and IceCube in the Minimal Ma Model}
\label{sec:IceCube-Ma}

As demonstrated in the previous section, the minimal Ma model equipped with an inert doublet scalar dark matter provides a natural framework for interpreting the LZ 248~keV nuclear recoil event through the endothermic inelastic scattering process
\begin{equation}
\eta_R + N \to \eta_I + N \, ,
\end{equation}
where $N$ denotes a target nucleus. 
The profile-likelihood analysis favors a mass splitting $\delta \equiv m_{\eta_I} - m_{\eta_R} \approx 360$--$380$~keV and a dark matter mass $m_{\eta_R} \sim \mathcal{O}(1)$~TeV, with the correct relic density achieved through co-annihilation among the nearly degenerate inert scalars. 

However, this parameter region is in severe tension with the null results of the IceCube neutrino observatory~\cite{IceCube:2025fcu}. 
In the minimal IDM, the inelastic scattering is mediated by the $Z$ boson, whose couplings to nucleons are completely fixed by the electroweak gauge structure. 
The effective vector coupling of the $Z$ boson to a nucleon $N$ is given by
\begin{equation}
f_N^Z = T_3^N - 2 Q_N \sin^2 \theta_W \, ,
\end{equation}
where $T_3^N$ is the third component of weak isospin, $Q_N$ is the electric charge. 
Using the measured value $\sin^2 \theta_W \approx 0.231$, we obtain
\begin{equation}
f_p^Z = \frac{1}{2} - 2 \sin^2 \theta_W \approx + 0.038 \, , \qquad f_n^Z = - \frac{1}{2} = - 0.500 \, .
\end{equation}
Although $f_p^Z$ is numerically small, it is non-vanishing. 
As a consequence, dark matter particles can scatter off protons in the Sun, lose kinetic energy, and become gravitationally captured. 
Once captured, the dark matter accumulates in the solar core and annihilates into SM particles---primarily $W^+ W^-$ and $Z Z$ at the TeV scale---producing high-energy neutrinos that are in principle detectable by the IceCube. 

A natural question is whether simply increasing the dark matter mass $m_{H^0}$ could suppress the capture rate. 
The spin-independent scattering cross section scales as
\begin{equation}
\sigma_N \propto \mu_{\chi N}^2 = \left( \frac{m_{\eta_R} m_N}{m_{\eta_R} + m_N} \right)^2 \, ,
\end{equation}
where $\mu_{\chi N}$ is the reduced mass of the dark matter--nucleon system. 
In the limit $m_{\eta_R} \gg m_N$, the reduced mass saturates at $\mu_{\chi N} \to m_N$, and the capture rate becomes essentially independent of $m_{\eta_R}$. 
Therefore, pushing the dark matter mass into the multi-TeV regime does not alleviate the IceCube bound. 
Recent phenomenological studies (see, e.g., ref.~\cite{Bose:2026ndd,Lee:2026xxh}) have confirmed that the minimal IDM parameter space favored by the LZ 248~keV event is essentially excluded by the current IceCube neutrino limits.

\subsection{Extension with a Hidden $U(1)_X$ Gauge Symmetry}
\label{sec:hiddenU1}

To resolve this tension while preserving the attractive features of the Ma model, we extend the gauge group of the theory by introducing a hidden $U(1)_X$ symmetry, 
%\begin{equation} G_{\text{SM}} \times U(1)_X \times Z_2.\end{equation}
where we assign a $U(1)_X$ charge $q_X$ to the inert doublet $\eta$, and introduce a new massive gauge boson $X_{\mu}$ associated with $U(1)_X$. 
The gauge kinetic Lagrangian contains a kinetic mixing term between the SM hypercharge field $B_{\mu}$ and the hidden gauge field $X_{\mu}$:
\begin{equation}
\mathcal{L} \supset - \frac{1}{4} B_{\mu \nu} B^{\mu \nu} - \frac{1}{4} X_{\mu \nu} X^{\mu \nu} - \frac{\epsilon}{2} B_{\mu \nu} X^{\mu \nu} + \frac{1}{2} M_X^2 X_{\mu} X^{\mu} \, ,
\end{equation}
where $B_{\mu \nu}$ and $X_{\mu \nu}$ are the field strength tensors of $B_{\mu}$ and $X_{\mu}$, respectively, and $\epsilon$ is the dimensionless kinetic mixing parameter. 
The $U(1)_X$ symmetry is assumed to be spontaneously broken by a hidden-sector scalar, generating the mass $M_X$ for the gauge boson $X$. 

After electroweak symmetry breaking, the neutral gauge boson mass matrix in the $(B, X)$ basis receives off-diagonal entries proportional to $\epsilon$. 
Diagonalizing the mass matrix in the limit $M_{Z'} \gg M_Z$, the physical mass eigenstates $Z$ and $Z'$ are related to the interaction eigenstates by a small mixing angle $\xi$, approximately given by
\begin{equation}
\xi \simeq \epsilon \tan \theta_W \left( 1 + \mathcal{O} \! \left( \frac{M_Z^2}{M_{Z'}^2} \right) \right) \, .
\end{equation}
Crucially, because the $Z'$ boson inherits a component of the SM $Z$ boson through this mixing, it also inherits the $SU(2)_L$ structure of the $Z$ couplings. 
In particular, the off-diagonal coupling $Z_{\mu} (\eta_R \partial^{\mu} \eta_I - \eta_I \partial^{\mu} \eta_R)$ present in the minimal IDM is transmitted to the $Z'$ boson. 
Therefore, the $Z'$-mediated inelastic scattering $\eta_R + N \to \eta_I + N$ remains operative, and the mechanism for explaining the LZ 248~keV event is fully preserved.

\subsection{Isospin-Violating Dark Matter and Cancellation of $f_p$}
\label{sec:IsoVio}

The decisive advantage of this extension is that the $Z'$ boson provides an additional contribution to the dark matter--nucleon scattering amplitude, which can interfere with the SM $Z$-mediated amplitude. 
At low momentum transfer, the effective coupling of dark matter to a nucleon $N$ becomes
\begin{equation}
f_N^{\text{eff}} = f_N^Z + \xi \frac{M_Z^2}{M_{Z'}^2} f_N^{Z'} \, ,
\end{equation}
where $f_N^{Z'}$ denotes the coupling of the $Z'$ boson to the nucleon, determined by the $U(1)_X$ charge $q_X$ and the quark content of the nucleon. 

We now impose the condition that the effective coupling to the proton vanishes,
\begin{equation}
f_p^{\text{eff}} = f_p^Z + \xi \frac{M_Z^2}{M_{Z'}^2} f_p^{Z'} = 0 \, ,
\end{equation}
in order to suppress the dark matter capture in the Sun, which is composed predominantly of hydrogen. 
This condition uniquely determines the required value of the kinetic mixing parameter:
\begin{equation}
\epsilon \simeq - \frac{f_p^Z}{f_p^{Z'}} \frac{M_{Z'}^2}{M_Z^2} \cot \theta_W \, .
\end{equation}
Since $f_p^Z \approx + 0.038$ is numerically small but non-zero, a moderate value of $\epsilon$ is sufficient to achieve the exact cancellation. 
For a benchmark $Z'$ mass $M_{Z'} \sim \mathcal{O}(1)$~TeV and $f_p^{Z'} \sim \mathcal{O}(1)$, one typically requires a range of $\epsilon \sim 10^{-3}$--$10^{-2}$, that is well within the reach of current and future precision experiments. 

On the other hand, the neutron coupling $f_n^Z = - 0.500$ is much larger in magnitude than the proton one. 
The $Z'$ contribution to $f_n^{\text{eff}}$ is suppressed by the factor $M_Z^2/M_{Z'}^2 \ll 1$ and does not significantly modify the neutron coupling, so that
\begin{equation}
f_n^{\text{eff}} \simeq f_n^Z = - 0.500 \, .
\end{equation}
The LZ detector employs liquid Xenon as the target material, whose dominant isotope $^{131}\text{Xe}$ has a large neutron number $N \approx 77$ and a proton number $Z = 54$. 
The coherent nuclear response is therefore dominated by the neutron coupling, and the dark matter can still scatter efficiently off Xenon nuclei, explaining the observed 248~keV recoil event. 

Quantitatively, the ratio of the scattering cross sections on Xenon and hydrogen is approximately
\begin{equation}
\frac{\sigma_{\text{Xe}}}{\sigma_{\text{H}}} \sim \frac{\left( N f_n^{\text{eff}} + Z f_p^{\text{eff}} \right)^2}{\left( f_p^{\text{eff}} \right)^2} \simeq \frac{(77 \times 0.5)^2}{0^2} \to \infty,
\end{equation}
demonstrating that the capture in the Sun is parametrically suppressed while the signal in LZ is unsuppressed. 

This mechanism transforms the minimal Ma model into a viable framework that simultaneously accommodates the LZ 248~keV signal, the observed dark matter relic density, and the stringent IceCube neutrino limits. 
Moreover, the required relation between $\epsilon$ and $M_{Z'}$ provides concrete, testable predictions for future $Z'$ searches at the LHC (via di-lepton resonance searches), at Belle II, and at low-energy fixed-target experiments, making this scenario falsifiable in the near future.

\subsection{Concrete model extended by the minimal radiative seesaw}
\label{sec:concretemodel}

\begin{figure}[t]
\begin{center}
\begin{tikzpicture}
\begin{feynman}[large]
%%% definition of each vertex (and auxiliary points) %%%
\vertex (a1) {\(L_L\)};
\vertex [right=1.8cm of a1] (b1);
\vertex [right=4.0cm of b1] (c1);
\vertex [right=1.5cm of c1] (d1) {\(L_L\)};
\vertex [right=2.0cm of b1] (m1);
\vertex[blob] [above=1.4cm of m1] (e1) {};
\diagram [medium] {
%%% define all lines and labels on propagators %%%
(a1) -- (b1) -- [insertion={[size=3pt]0.5}, edge label'=\(N_R~~~~~~~~~~~N_R\)] (c1) -- (d1),
(b1) -- [scalar, quarter left, edge label=\(\eta\), looseness=0.9] (e1),
(c1) -- [scalar, quarter right, edge label'=\(\eta\), looseness=0.9] (e1),
};
\end{feynman}
\end{tikzpicture}
\end{center}
\vspace{-0.7cm}
\caption{Extension of the Ma model, where there are several possibilities how to construct the tiny $\lambda_5$ indicated by the gray blob. }
\label{fig:ma_ext}
\end{figure}
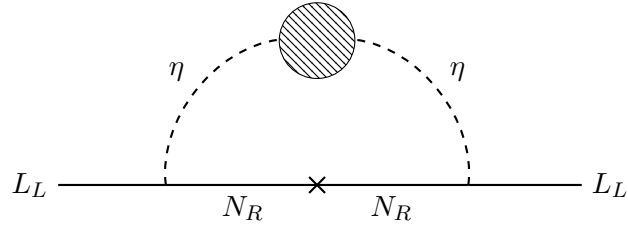

Since our DM candidate must have nonzero charge $q_X$ under the gauged $U(1)_X$ symmetry, $\lambda_5$ cannot be generated at tree level unless $H$ is also charged under the $U(1)_X$. 
This fact would naturally lead us to a scenario that $\lambda_5$ has to be induced at loop level in Fig.~\ref{fig:ma_ext}. 
%%%%%%%%%%%%%%%%%%%
%%%%%%%%%%%%%%%%%%%
This is a reasonable extension since $\lambda_5$ is expected to be tiny $\sim {\cal O} (10^{-5})$. 
As one of the concrete scenarios to realize one-loop induced $\lambda_5$, we propose the following model set up.\footnote{Several models can be found in Refs.~\cite{Aoki:2013gzs,Cai:2017jrq,CarcamoHernandez:2026pis} along this line of ideas.} 
We introduce two inert singlet scalars $\chi$ and $\chi'$ with $- 2 q_X$ and $+ q_X$ under the $U(1)_X$, respectively. 
Furthermore, we need to assign $N_R$ to be $q_X$ to allow $Y$ in Eq.~\eqref{eq:yukawa}. 
In addition, a new $Z_2$ odd is imposed to these two new particles, while $Z_2$ even is assigned for the other particles. 
Note that the original $Z_2$ in Ma model is not needed since the hidden $U(1)_X$ is replaced by the $Z_2$. 
In order to spontaneous symmetry breaking of $U(1)_X$, we introduce another singlet scalar $\varphi$ with $- 2 q_X$, where $\langle \varphi \rangle \equiv v' / \sqrt{2}$. 
$\varphi$ plays a crucial role in providing nonzero masses for $N_R$, $\chi$, and $\chi'$, as well as $Z'$ boson mass. 
We summarize our particle contents and their assignments in Table~\ref{hu1model}. 
\begin{widetext}
\begin{center} 
\begin{table}[t]
%%%%%%%%%%%
\begin{tabular}{|c||c|c|c|}\hline\hline
Fermions   & ~$L_L$~          & ~$e_R$~  & ~$N_R$~  \\ \hline
$SU(2)_L$  & $\bm{2}$         & $\bm{1}$ & $\bm{1}$ \\ \hline
$U(1)_Y$   & $- \frac{1}{2}$  & $-1$     & $0$      \\ \hline
$U(1)_{X}$ & $0$              & $0$      & $q_X$    \\ \hline
$Z_2$      & $+$              & $+$      & $+$      \\ \hline
\end{tabular}
%%%%%%%%%%%
\begin{tabular}{|c||c|c|c|c|}\hline\hline
Bosons     & ~$H$~         & ~$\varphi$~ & ~$\chi$~ & ~$\chi'$~ \\ \hline
$SU(2)_L$  & $\bm{2}$      & $\bm{1}$    & $\bm{1}$ & $\bm{1}$  \\ \hline 
$U(1)_Y$   & $\frac{1}{2}$ & $0$         & $0$      & $0$       \\ \hline
$U(1)_{X}$ & $0$           & $- 2q_X$    & $- 2q_X$ & $q_X$     \\ \hline
$Z_2$      & $+$           & $+$         & $-$      & $-$       \\ \hline
\end{tabular}
%%%%%%%%%%%
\caption{Field contents of fermions and bosons
and their charge assignments under $SU(2)_L \times U(1)_Y \times U(1)_{X} \times Z_2$, where flavor indices are abbreviated and $q_X \neq 0$.}
\label{hu1model}
\end{table}
\end{center}
\end{widetext}

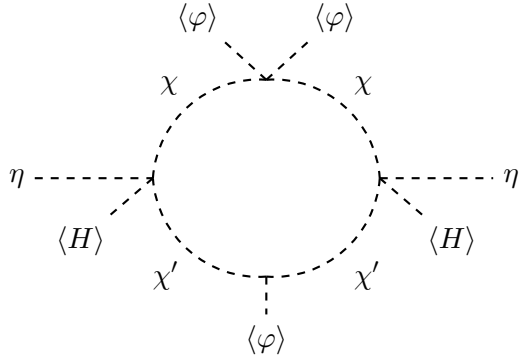
\begin{figure}[t]
\begin{center}
\begin{tikzpicture}
\begin{feynman}[large]
%%% definition of each vertex (and auxiliary points) %%%
\vertex (a2) {\(\eta\)};
\vertex [right=1.8cm of a2] (b2);
\vertex [right=3.0cm of b2] (c2);
\vertex [right=1.5cm of c2] (d2) {\(\eta\)};
\vertex [right=1.5cm of b2] (m2);
\vertex [above=1.3cm of m2] (e2);
\vertex [below=1.3cm of m2] (f2);
\vertex [below left=0.7cm of b2] (hl2) {\(\langle H \rangle\)};
\vertex [below right=0.7cm of c2] (hr2) {\(\langle H \rangle\)};
\vertex [above left=0.7cm of e2] (vpal2) {\(\langle \varphi \rangle\)};
\vertex [above right=0.7cm of e2] (vpar2) {\(\langle \varphi \rangle\)};
\vertex [below=0.5cm of f2] (vpl2) {\(\langle \varphi \rangle\)};
\diagram [medium] {
%%% define all lines and labels on propagators %%%
(a2) -- [scalar] (b2) -- [scalar, quarter left, edge label=\(\chi\)] (e2) -- [scalar, quarter left, edge label=\(\chi\)] (c2) -- [scalar] (d2),
(c2) -- [scalar, quarter left, edge label=\(\chi'\)] (f2) -- [scalar, quarter left, edge label=\(\chi'\)] (b2),
(b2) -- [scalar] (hl2),
(c2) -- [scalar] (hr2),
(vpal2) -- [scalar] (e2) -- [scalar] (vpar2),
(f2) -- [scalar] (vpl2),
};
\end{feynman}
\end{tikzpicture}
\end{center}
\vspace{-0.7cm}
\caption{An example of the blob part in Fig.~\ref{fig:ma_ext}. 
$\langle H \rangle$ and $\langle \varphi \rangle$ are corresponding VEVs. }
\label{fig:lam5-1lp}
\end{figure}

Under these symmetries, only the Majorana mass term  is extended as follows:
\begin{equation}
- \frac{1}{2} \sum_i  y_{\varphi_i} \varphi \overline{N^C_{R_i}} N_{R_i} + \mathrm{h.c.} \, ,
\label{eq:yukawaU(1)H}
\end{equation}
where $M_i$ can be obtained by $y_{\varphi_i} v' / \sqrt{2}$ after the spontaneous symmetry breaking of $\varphi$. 
The relevant non-trivial terms are added in the original Higgs potential as follows:
\begin{align}
+ \frac{1}{2} \mu_{\chi'} \varphi \chi'^2 + \lambda_{\varphi \chi} (\varphi^*)^2 \chi^2 + \lambda_0 (H^{\dag} \eta) \chi \chi' + \mathrm{h.c.} \, .
\label{eq:extpot}
\end{align}
From these new terms, we can generate $\lambda_5$ at the one-loop diagram shown in Fig.~\ref{fig:lam5-1lp}, and its form is given by~\cite{Aoki:2013gzs}
\begin{align}
\lambda_5 \simeq \frac{\lambda_0^2}{(4 \pi)^2} \left( \frac{m_{\chi_R}^2}{m_{\chi_R}^2 - m^2_{\chi'}} \ln \left[ \frac{m_{\chi_R}^2}{m_{\chi'}^2} \right] - \frac{m_{\chi_I}^2}{m_{\chi_I}^2 - m^2_{\chi'}} \ln \left[ \frac{m_{\chi_I}^2}{m_{\chi'}^2} \right] \right) \, ,
\end{align}
where $m_{\chi_R}$, $m_{\chi_I}$, and $m_{\chi'}$ are respectively the mass eigenvalues for $\chi_R$, $\chi_I$, and $\chi'$. 
%%%%%%%%%%%%%%%%%%%
%%%%%%%%%%%%%%%%%%%
The mass difference between $\chi_R$ and $\chi_I$ is parameterized by $\lambda_{\varphi \chi}$. 
Then, the desired value of $\lambda_5$ will be realized by certain values of $\lambda_0$, $m_{\chi_{R, I}}$ and $m_{\chi'}$, and we can discuss rich phenomenology including these hidden particles. 
The details of such investigations are left for future works.

\subsection{Discussion: Phenomenological Constraints and Interpretation}
\label{sec:discussion}

In this section, we discuss the implications of our benchmark points for collider searches and cosmology, and clarify the scope of our interpretation of the LZ high-recoil event. 
While the tree-level elastic scattering and relic density calculations establish the viability of the parameter space, a complete assessment requires considering additional constraints. 

\subsubsection*{Collider Constraints}

The scalar spectrum in our benchmark points is characterized by a compressed mass hierarchy, with the charged scalar mass splitting $\Delta^{\pm} = m_{\eta^{\pm}} - m_{\eta_R}$ ranging from approximately $6.3$ to $21.2$~GeV.
In this regime, the dominant decay mode of the charged scalar is the three-body decay $\eta^{\pm} \to \eta_{R, I} W^*$, mediated by an off-shell $W$ boson. 
This soft decay product poses a significant challenge for standard LHC search strategies, which typically rely on high-$p_T$ leptons or large missing transverse energy. 

For a subset of our points, the mass gap is large enough to kinematically open the two-body decay channel $\eta^{\pm} \to \ell^{\pm} N_i$, provided the Yukawa couplings are non-negligible. 
However, the opening of this channel does not automatically guarantee dominance over the gauge-mediated mode, nor does it imply immediate detectability. 
A rigorous exclusion limit would require a dedicated simulation incorporating production cross-sections, branching fractions, and the specific acceptance of soft-lepton or disappearing-track searches at ATLAS and CMS. 
Therefore, while our points are subject to collider bounds, we do not make a quantitative claim of exclusion in this work, noting that the compressed spectrum naturally suppresses the visibility in standard search channels. 

\subsubsection*{Cosmological Constraints: BBN and CMB}

The small mass splitting $\delta \simeq 360 $ ~keV between the neutral scalars implies that the excited state $\eta_I$ is long-lived. 
The dominant decay channel is $\eta_I \to \eta_R \nu \bar{\nu}$ via an off-shell $Z$ boson. 
For the benchmark points considered, the lifetime is estimated to be $\tau_{\eta_I} \sim 4\text{--}6$ days. 

A lifetime in this range raises concerns regarding BBN and the Cosmic Microwave Background (CMB). 
However, the phenomenological impact is determined not just by the lifetime but by the energy of the injected particles. 
The kinetic energy available to the decay products is bounded by $\delta \simeq 360 \kev$, which is well below the thresholds for electron-positron pair production ($2 m_e \simeq 1.02$~MeV) and hadronic interactions. 
Consequently, the decay injects only soft neutrinos into the thermal bath. 

We estimate the energy injection into the electromagnetic sector to be negligible. 
Assigning the full reference dark matter abundance to the excited state $\eta_I$ (a conservative overestimate), the ratio of injected energy to radiation density is well below the critical values required to alter the neutron-to-proton ratio or distort the CMB blackbody spectrum. 
The corresponding contribution to the effective number of neutrino species, $\Delta N_{\text{eff}}$, is estimated to be $\lesssim 1.5 \times 10^{-9}$, which is orders of magnitude below current observational sensitivities. 
Thus, the long-lived nature of $\eta_I$ does not pose a threat to standard cosmology in this specific parameter region. 

\subsubsection*{Fermi-LAT Constraints}

The annihilation cross section $\langle \sigma v \rangle$ is directly constrained by indirect detection experiments, particularly Fermi-LAT observations of gamma-ray emission from dwarf spheroidal galaxies (dSphs) and the Galactic Center \cite{Fermi-LAT:2016uux}. 

The Fermi-LAT has placed stringent upper limits on the velocity-averaged annihilation cross section $\langle \sigma v \rangle$ by searching for gamma-ray signals from DM annihilation in dSphs. 
For the mass range relevant to the Ma model ($m_{\eta_R} \sim 100~\text{GeV} - 1~\text{TeV}$), the constraints on the $W^+ W^-$ and $b \bar{b}$ final states are particularly severe. 
The canonical thermal relic cross section $\langle \sigma v \rangle \sim 3 \times 10^{-26}~\text{cm}^3 / \text{s}$ is in tension with or marginally excluded by the Fermi-LAT limits around $m_{\eta_R} \sim 100~\text{GeV}$. 
Moreover, combined analyses incorporating data from Fermi-LAT, H.E.S.S., MAGIC, VERITAS, and HAWC have further strengthened these constraints, especially in the TeV mass regime where ground-based Cherenkov telescopes achieve their highest sensitivity. 

Interpreting recent LZ observation within the Ma model/IDM framework requires a small mass splitting $\Delta m \sim \mathcal{O} (100) \kev$ between the DM candidate and its slightly heavier partner (e.g., $\eta_R \to \eta_I$), enabling endothermic scattering off Xenon nuclei. 
%However, a significant tension arises from the observed relic density $\Omega_{\text{DM}} h^2 \approx 0.12$ for $m_{\text{DM}} \gtrsim 500 \gev$, efficient annihilation into electroweak gauge bosons is necessary, and from the same efficient annihilation that sets the relic density in the early Universe predicts a present-day $\langle \sigma v \rangle$ that may exceed Fermi-LAT upper limits, leading to exclusion of the parameter space. 
%However, a significant tension arises for $m_{\text{DM}} \gtrsim 500 \gev$. The observed relic density, $\Omega_{\text{DM}} h^2 \approx 0.12$, requires efficient annihilation into electroweak gauge bosons, but this same annihilation process can lead to a present-day $\langle \sigma v \rangle$ exceeding the Fermi-LAT upper limits, thereby excluding the corresponding parameter space.
However, for $m_{\text{DM}} \gtrsim 500 \gev$, a significant tension emerges between the relic-density requirement and indirect-detection constraints. 
Reproducing the observed relic density, $\Omega_{\text{DM}} h^2 \approx 0.12$, requires efficient annihilation into electroweak gauge bosons, which can in turn yield a present-day $\langle \sigma v \rangle$ above the Fermi-LAT upper limits and exclude the corresponding parameter space. 
Several mechanisms have been proposed to reconcile the LZ event interpretation with Fermi-LAT constraints in Refs.~\cite{Garcia-Cely:2015quu,Eiteneuer:2017hoh,Bandyopadhyay:2026gjw}. 
While Fermi-LAT constraints from dSph gamma-ray observations pose a significant challenge for the Ma model/IDM interpretations of the LZ 248~keV event, viable parameter space remains accessible through careful model building. 
The key strategy involves decoupling the early-Universe annihilation dynamics from present-day observables, typically via mass degeneracy and co-annihilation mechanisms. 
Future observations by the Cherenkov Telescope Array and next-generation direct detection experiments will provide decisive tests of these scenarios.

\section{Conclusions}
\label{sec:conclusion}

We have proposed a comprehensive and highly motivated framework to interpret the recently reported 248~keV high-energy nuclear recoil event in the LZ experiment within the minimal radiative seesaw model. 
By identifying the DM candidate as the neutral component of the inert scalar doublet, we demonstrated that a sub-MeV mass splitting ($\delta \simeq 360 \kev$) between the real and imaginary neutral components naturally explains the anomaly. 
This splitting kinematically forbids low-energy elastic scattering, thereby evading stringent direct detection bounds, while permitting rare inelastic scattering events driven by the high-velocity tail of the dark matter halo. 

A critical advantage of our setup over the pure IDM is the thermal history of the dark sector. 
The inclusion of right-handed Majorana fermions enables efficient co-annihilation processes, which significantly enhance the relic density and allow the dark matter mass to reach up to $\sim {\cal O} (1)$~TeV 
%$\sim 1.5$ TeV 
while keeping the Higgs portal couplings small. 
Through an extensive numerical scan, we identified 34 viable parameter points satisfying all theoretical and experimental constraints, including neutrino oscillation data and cLFVs. 
For these points, the dark matter mass spans $\mR \in [691, 1479]$~GeV, and we explicitly calculated the expected signal yields in the LZ working-search selection, finding $N_{\rm WS} \simeq 0.2$--$4.1$ events for our benchmark scenarios, which quantitatively matches the scale of the observed rare high-energy signal.

To resolve the severe tension between the minimal model and the IceCube neutrino observatory limits regarding solar capture, we extended the framework by introducing a hidden $U(1)_X$ gauge symmetry that naturally leads us to tiny $\lambda_5$ coupling at the one-loop level. 
Through kinetic mixing between the SM $Z$ boson and the new $Z'$ boson, we achieved an isospin-violating dark matter scenario. 
This precisely cancels the effective dark matter-proton coupling, suppressing capture in the Sun, while preserving the dark matter-neutron coupling necessary to generate the coherent signal in the Xenon-based LZ detector. 

Finally, we verified that the proposed parameter space safely satisfies cosmological constraints from BBN and the CMB, as the soft neutrino injection from the long-lived excited state poses no threat to standard cosmology. 
The compressed scalar spectrum and the hidden gauge sector provide rich phenomenological footprints, offering concrete and testable predictions, such as the required kinetic mixing $\epsilon \sim \mathcal{O} (10^{-3}$--$10^{-2})$, for future collider searches at the LHC and Belle II, as well as next-generation direct and indirect detection experiments.

%%%%%%%%%%%%%%%%%%%%%%%%%%%%%%%%%%%
\section*{Acknowledgments}
H.O. is supported by Zhongyuan Talent (Talent Recruitment Series) Foreign Experts Project. 
Y.S. is supported by Natural Science Foundation of China under grant No. W2433006. 
%%%%%%%%%%%%%%%%%%%%%%%%%%%%%%%%%%%
 
\bibliography{ref}

\end{document}